# Magnetization reversal in exchange-spring bilayer system under circularly polarized microwave field

R. Tayade[1], A. Pasko[1], C. Serpico[2], F. Mazaleyrat[1] and M. LoBue[1]

[1]SATIE, ENS Cachan, CNRS, UniverSud, Cachan, 94230, France
[2]Dipartimento di Ingegneria Elettrica, University di Napoli, Napoli, Italy

*Microwave assisted magnetization reversal are studied in the bulk bilayer exchange coupled system. We investigate the nonlinear magnetization reversal dynamics in a perpendicular exchange spring media using Landau-Lifshitz equation. In the limit of the infinite thickness of the system, the propagation field leads the reversal of the system. The reduction of the switching field and the magnetization profile in the extended system are studied numerically. The possibility to study the dynamics analytically is discussed and an approximation where two P-modes are coupled by an interaction field is presented. The ansatz used for the interaction field is validated by comparison with the numerical results. This approach is shown to be equivalent to two exchange coupled macrospins.*

## I. Introduction

With the tremendous demand in the electronic storage technology, new strategies are being continuously tested to overcome the so called "recording trilemma" [1]. This fast-paced developing technology requires understanding of the fundamental dynamical magnetization reversal processes [2] [3]. One of the current challenges is to reduce the writing field in the recording devices as the use of high perpendicular-anisotropy materials requires very intense applied fields for switching the media grains.

One of the projected solutions is to use the heat-assisted magnetic recording (HAMR) or thermally assisted recording (TAR) technique [4]. Thirion et al. [5] proposed a method to overcome high fields required to reverse the spins in the high density media without compromising the thermal stability of the devices by introducing a radio-frequency (RF) field along with a constant applied field [6]. Suess et al. [7] reviewed the use of exchange spring media for magnetic recording and moreover they showed that it is possible for perpendicular recording media to achieve the required thermal stability at very high densities.

In this work we focus on the application of microwave assisted switching to perpendicular exchange spring media (PESM). Many works have been devoted to show the reduction in the switching field in PESM compared to single phase high anisotropic media [8], [9], [10]. However, there is very limited literature to explain analytically the mechanisms of switching and reversal in this class of materials [11]. The analytical description of magnetic behavior in PESM was studied by Asti et al. [12] using the static micromagnetic approach. The magnetization dynamics was analytically studied by Bertotti et al. [13], [14] for uniaxial anisotropic media. This work was limited to uniformly magnetized systems.

The challenge represented by the non linear nature of Landau Lifshitz (LL) equation, and by the non-uniformity of the problem, due to the interface between the hard and soft magnetic phases, makes a pure analytical solution hardly achievable. However, here we show that the numerical solution of the problem makes possible, under rather general condition, to describe the dynamics using the same analytical results obtained for the decoupled layers with an additional interaction field mimicking the effect of the interface induced nonuniformity.

## II. Model

We consider a system composed of two layers lying on the *xy* plane, one magnetically hard and the other soft. Both layers exhibit perpendicular uniaxial anisotropy directed along *z*-axis. The system is infinite in the *xy* plane and due to this geometry the magnetization is uniform along the plane. Thus the problem reduces to work out the distribution of magnetization vector **M**= **M**(z) with respect to *z*, is the local magnetization vector. Calling $z_0$ the coordinate of the interface between the soft and the hard layer we have that |**M**(z)| = $M_s(z)$. The in-plane component of the magnetization is given by $M_\perp = \sqrt{M_x^2 + M_y^2}$. The magnetocrystalline anisotropy is $K(z) = K_1$ for $z < z_0$, and $K(z) = K_2$ otherwise. The shape and magnetocrystalline anisotropies can be merged in a single effective anisotropy constant, $K_u(z) = K(z) - \mu_0 M_s^2(z)/2$. The two phases are assumed to be exchange coupled at the interface.

The magnetization dynamics are given by LL equation,

$$\frac{(1+\alpha^2)}{\gamma}\frac{\partial \mathbf{M}}{\partial t} = -\mathbf{M}\times\mathbf{H}_{eff} - \frac{\alpha}{M_s(z)}\mathbf{M}\times(\mathbf{M}\times\mathbf{H}_{eff}) \cdot \quad (1)$$

where γ is the gyromagnetic ratio associated with electron spin, α is the damping constant, and **H**$_{eff}$ is the effective field,

$$\mathbf{H}_{eff} = \mathbf{H}_a(t) + \frac{2K_u(z)}{\mu_0 M_s(z)}\mathbf{e_z} + l_{ex}^2\frac{\partial^2 \mathbf{M}}{\partial z^2}, \cdot \quad (2)$$

where $l_{ex} = (2A/\mu_0 M_s^2(z))^{1/2}$ is the characteristic exchange length.

**H**$_a$(t) is the total applied field which is the sum of a DC magnetic field, **H**$_{az}$ applied perpendicular to the plane of the media and a circularly polarized microwave field of magnitude $H_{a\perp}$ rotating with angular frequency $\omega = 2\pi f$, applied in the plane.

$$\mathbf{H}_a(t) = H_{az}\mathbf{e}_z + H_{a\perp}(\cos\omega t\, \mathbf{e}_x + \sin\omega t\, \mathbf{e}_y). \quad (3)$$

It is known [12] that there are two fields controlling the process of magnetic reversal in spring systems, the nucleation field $H_N$ and propagation field $H_P$. For exchange coupled multilayer systems it has been shown that the leading field in the reversal process, for thin systems is $H_N$ and $H_P$ in the limit of infinite thickness of the layers. In this case, the analytical expression for the propagation field in static conditions is [12],

$$H_P = \frac{2K_{u1}}{\mu_0 M_s}\left[1 - \sqrt{1 - \frac{K_{u2}}{K_{u1}}}\right]. \quad (4)$$

To study numerically the dynamical reversal of this system we have developed a Fortran code solving LL equation using the finite difference technique [15]. Here we focus our attention on a thick system where the propagation field leads the reversal process. Actually the infinite thickness limit is easily reached when the total width of the system is few tens of its domain wall.

The switching process is studied keeping $H_{a\perp}$ and $\omega$ constant and by slowly changing the DC field between $H_{az} = -1.6 \times 10^6$ A/m and $H_{az} = 1.6 \times 10^6$ A/m. The effect of eddy currents on the uniformity of the applied field is neglected.

We solve the equation of motion for the bilayer system using the micromagnetic parameters of a widely used PESM, FePt/Fe system. The parameters used are as follows: exchange constant $A_1=A_2=10^{-11}$ J/m, anisotropy constants $K_1 = 2 \times 10^6$ J/m$^3$, $K_2 = 0.48 \times 10^5$ J/m$^3$. We performed calculations assuming same saturation for both layers $M_s = 1.55 \times 10^6$ A/m (this overestimates $M_s$ for hard layer but allows a simpler analytical formulation in term of reduced units). Damping parameter is taken to be $\alpha = 0.01$. Soft and hard layers have equal thicknesses of 100 nm. The magnitude of the in-plane microwave field is kept constant, $H_{a\perp} = 8 \times 10^4$ A/m. Simulations are performed for several microwave frequencies ranging 1 GHz $\leq f \leq$ 20 GHz.

### III. RESULTS AND DISCUSSION

A first and worth discussing result of the simulations is that, in a frequency interval up to more than 10 GHz, the average magnetization of the system, taken along $z$, $<M_z>$, behaves like a P-mode. A P-mode is a stable solution of the dynamic equation for the system subjected to the microwave field, which results in uniform precession of the magnetization around anisotropy axis, in synchronization with microwave field. Due to non-uniform magnetization in the bilayer system, the stable solution is given by the average magnetization $<M_z>$, which is constant in time forming an angle $\theta$ with respect to $z$ axis whereas $<\mathbf{M}_\perp>$ is rotating with angular frequency $\omega$, and with a phase lag $\varphi$ with respect to the rotating component of the applied field $\mathbf{H}_{a\perp}$. We call this stationary state a *global P-mode*. In Fig.1 we plot the normalized average $z$-component of the magnetization against the DC magnetic field showing the ascending branch of a switching loop with $-1.6 \times 10^6 \leq H_a \leq 1.6 \times 10^6$ A/m. The P-mode behavior is apparent up to $f = 10$ GHz. Beyond this limit, the average magnetization shows oscillations whose behavior is much more similar to a quasi periodic mode (Q-mode, using terminology from [16] [17]. Another feature apparent in Fig. 1 is the switching field reduction as a function of the microwave frequency (see inset in Fig.1). This microwave assisted switching is similar to the one reported in [13] for uniformly magnetized systems. It is worth noting that 6 GHz curve is switching at a very small positive DC field. This results opens a possibility to investigate zero field directional field microwave assisted switching.

A deeper analysis of the simulation results gives us the opportunity to go beyond the global P-mode concept. Fig.2 shows the angle $\theta(z)$ between perpendicular component of magnetization $\mathbf{M}$ and $\mathbf{H}_{az}$ as a function of $z$ for different values of $H_{az}$. The system is subjected to a constant in-plane circularly polarized field with $f = 6$ GHz. The figure shows two uniform regions separated by a thin transition region. The thickness of the latter stays constant (~30 nm) all along the magnetization curve and its contribution to the average magnetization can be assumed to be negligible. A plot of the lag angle $\varphi(z)$ between $\mathbf{M}_\perp(z)$ and $\mathbf{H}_{a\perp}$ gives very similar results. Thus we can say that the global P-mode is nothing but the average of two individual P-modes, corresponding respectively to the soft and the hard layer.

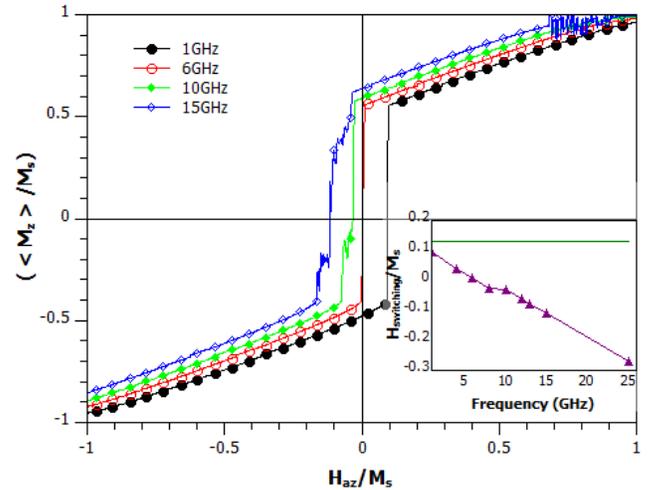

Fig.1. Spatial average of $M_z$ as a function of frequency. Only the branches of the loop when DC field is varied from $-1.6 \times 10^6$ to $1.6 \times 10^6$ A/m, is shown here for frequencies of 1, 6, 10 and 15 GHz. Variation of switching field in response to the microwave frequency (line with triangles) is shown in the inset with the solid straight line corresponding to the switching field in absence of microwave field.

Treating with P-mode solutions, it is possible to greatly simplify expressions by changing the system frame from the laboratory frame to a frame in which the system is rotating around $z$ axis with the same angular frequency $\omega$ of the microwave field. In this frame, the applied field becomes constant and is given by $\mathbf{H}_a = H_{az}\mathbf{e}_z + H_{ax}\mathbf{e}_x$ and the LL equation (1) is written as,

$$\frac{(1+\alpha^2)}{\gamma}\frac{\partial \mathbf{M}}{\partial t} = -\mathbf{M} \times \left[\mathbf{H}_{eff} - (1+\alpha^2)\frac{\omega}{\gamma}\mathbf{e}_z\right] - \frac{\alpha}{M_s(z)}\mathbf{M} \times$$
$$\times (\mathbf{M} \times \mathbf{H}_{eff}),$$
(5)

The magnetization $\mathbf{M}$ is fully defined by its angle $\theta$ with respect to z-axis and the angle $\varphi$ with respect to x (i.e. to $\mathbf{H}_{a\perp}$). From [16], and [17] we know that a P-mode maps, for given $\omega$, the values of $M_z$ and of the lag angle $\varphi$ (namely of $M_x$, and $M_y$) onto the values of the DC field $H_z$, and on the in-plane rotating field amplitude $H_\perp$ through the following expressions:

$$H_z = \frac{\omega}{\gamma} - \left[\nu_0 + \frac{2K_u}{\mu_0 M_s}\right]M_z \quad (6)$$

$$H_\perp = \sqrt{\left(\nu_0^2 + \frac{\alpha^2 \omega^2}{\gamma^2 M_s^2}\right)(M_s^2 - M_z^2)} \quad (7)$$

$$\nu_0 = \frac{\alpha \omega}{\gamma M_s}\frac{M_x}{M_y} \quad (8)$$

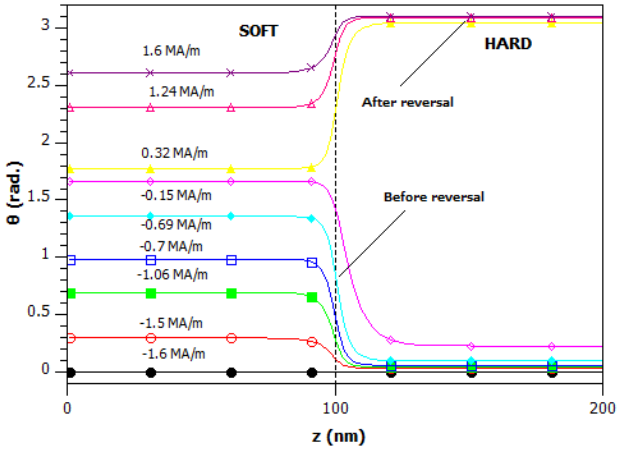

Fig.2 Spatial distribution of θ(z) for various DC field values at AC field frequency 6GHz. From z=0-100 represents the soft phase and z=100-200 represents hard phase. The black circles at θ=π represents the initial state of the system and the down triangle represents the final state of the system after complete switching.

From Eq. (6), (7), and (8), we can easily verify that the field we are actually applying, $\mathbf{H}_a$ would not produce the observed P-mode when applied to the uncoupled layers. From this we can suppose the observed P-mode to be generated by a field $\mathbf{H}_i$ (subscript $i = 1, 2$ refer to the P-mode we observe in $i$ layer) being the sum of the applied field and of an interaction field $\mathbf{H}_{j,\ exc}$ representing the interface exchange coupling with the other layer,

$$\mathbf{H}_i = \mathbf{H}_a + \mathbf{H}_{j,exc}. \quad (9)$$

We use a standard *ansatz* for the exchange field (see [11]) with $\mathbf{H}_{j,exc} = C\ \mathbf{M}_j$, where $C$ is a constant to be determined. Assuming that,

$$\mathbf{M}(z,t) = \begin{cases} \mathbf{M}_1(t) & \text{if } 0 \leq z \leq z_1 \\ \mathbf{M}_2(t) & \text{if } z_2 \leq z \leq L \end{cases} \quad (10)$$

where, $L$ is the total thickness of the system and $z_1 \leq z \leq z_2$, is the region where the transition layer is located. We define $|\mathbf{M}_{1,2}(t)| = M_{s1,2}$. The interaction due to the layer is taken into account by an energy term $-\mu_0 C \mathbf{M}_1 \cdot \mathbf{M}_2$, so that the effective field in the two spatially uniform layers is given by

$$\mathbf{H}_{eff\ 1,2} = \mathbf{H}_a + \left[\frac{2K_{1,2}}{\mu_0 M_{s1,2}^2} - 1\right]M_{z1,2}\hat{e}_z + C\mathbf{M}_{2,1} \quad (11)$$

The P-mode in the two coupled layers is given by the equations

$$-\mathbf{M}_{1,2} \times \left[\mathbf{H}_{eff\ 1,2} - (1+\alpha^2)\frac{\omega}{\gamma}\mathbf{e}_z\right] - \frac{\alpha}{M_{s1,2}}\mathbf{M}_{1,2} \times$$
$$\times (\mathbf{M}_{1,2} \times \mathbf{H}_{eff\ 1,2}) = 0$$
(12)

The four unknowns of the problem, specifically the three components of $\mathbf{H}_i$ and $C$, are fully determined by equations (6), (7), (9) and by the constraint, $|\mathbf{H}_{j,exc}| = C\ M_s$.

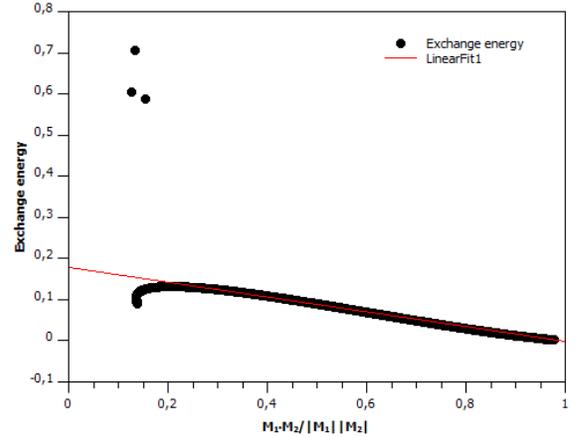

Fig.3 Representation of exchange energy versus the angle between $\mathbf{M}_1$ and $\mathbf{M}_2$ defined in Eq.(10).

Using our numerical results we calculated, $C = 0.1$. On the other hand, we can also estimate $C$ by calculating the exchange energy of the bilayer system. The exchange energy term is calculated by integrating $(\nabla \mathbf{M})^2$ along the sample thickness. This integration includes the transition region, which, of course, will give the main contribution to the total exchange energy of the system. Exchange energy is plotted against the angle between the magnetizations of the two uniform regions (i.e. the two P-modes), respectively $\mathbf{M}_1$ and $\mathbf{M}_2$, Fig.3. The slope of the curve gives the exchange coefficient $C$. Linear fitting of the curve is shown by the red line in the figure.

The points in the figure which do not follow the linear behavior correspond to the region around switching field.

This approach is equivalent to describe the system as two exchange coupled macrospin with different uniaxial anisotropy. This model is numerically solved under the same parametric conditions used for the bilayer system with the exchange constant derived from the solutions of bilayer system. In Fig.4 we show that $<M_z>(H_{az})$ loops for the extended bilayer and of the macrospin system are in very good agreement.

Further investigations will be devoted to apply the bifurcation analysis presented in [13] to the case of two coupled P-modes. In this case, the phase portrait analysis discussed in [13] can be used to describe the dynamics of each single P-mode substituting the field of Eq. (9) to the standard applied field.

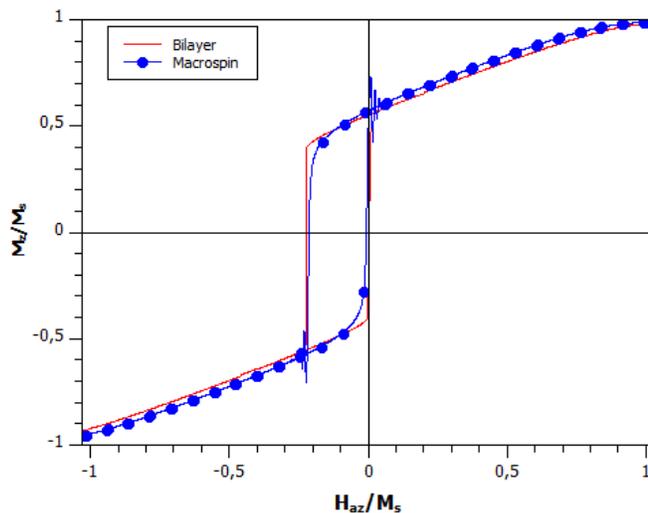

Fig.4 Average magnetization vs DC field. The loop in red (solid line) is the magnetization behavior for bilayer system obtained from the numerical simulations and the loop in blue (with filled circles) is the magnetization behavior from the macrospin magnetization model.

ACKNOWLEDGMENT

The authors wish to thank Giorgio Bertotti for many helpful suggestions. This work was funded by the ANR project MICROMANIP and by the *Fondation de Cooperation Scientifique* Campus Paris Saclay as a part of the foreign guest project of DIGITEO N◦ 2012-XXD. This work was partially supported by MIUR-PRIN Project No.2010ECA8P3 "DyNanoMag."